\documentclass[pdflatex,twocolumn,sn-mathphys-num]{sn-jnl}% Math and Physical Sciences Numbered Reference Style 
\usepackage{graphicx}%
\usepackage{multirow}%
\usepackage{amsmath,amssymb,amsfonts}%
\usepackage{amsthm}%
\usepackage{mathrsfs}%
\usepackage[title]{appendix}%
\usepackage{xcolor}%
\usepackage{textcomp}%
\usepackage{manyfoot}%
\usepackage{booktabs}%
\usepackage{algorithm}%
\usepackage{algorithmicx}%
\usepackage{algpseudocode}%
\usepackage{listings}%
\usepackage{upgreek}
\usepackage{subcaption}
\DeclareUnicodeCharacter{0308}{HERE!HERE!}

\DeclareCaptionLabelFormat{nocaption}{}
\begin{document}

\title[Large tunable kinetic inductance in a twisted graphene superconductor]{Large tunable kinetic inductance in a twisted graphene superconductor}

\author*[1,2]{\fnm{Rounak} \sur{Jha}}\email{rounak.jha@epfl.ch}
\author[1]{\fnm{Martin} \sur{Endres}}
\author[3]{\fnm{Kenji} \sur{Watanabe}}
\author[4]{\fnm{Takashi} \sur{Taniguchi}}
\author[2]{\fnm{Mitali} \sur{Banerjee}}
\author[1,5]{\fnm{Christian} \sur{Sch\"{o}nenberger}}
\author*[1]{\fnm{Paritosh} \sur{Karnatak}}\email{paritosh.karnatak@unibas.ch}

\affil*[1]{\orgdiv{Department of Physics}, \orgname{University of Basel}, \orgaddress{\city{CH-4056 Basel}, \country{Switzerland}}}

\affil[2]{\orgdiv{Laboratory of Quantum Physics (LQP)}, \orgname{\'{E}cole Polytechnique F\'{e}d\'{e}rale de Lausanne (EPFL)}, \orgaddress{\city{CH-1015 Lausanne}, \country{Switzerland}}}

\affil[3]{\orgdiv{Research Center for Functional Materials}, \orgname{National Institute for Material Science}, \orgaddress{\street{Street}, \city{1-1 Namiki}, \state{Tsukuba 305-0044}, \country{Japan}}}

\affil[4]{\orgdiv{International Center for Materials Nanoarchitectonics}, \orgname{National Institute for Material Science}, \orgaddress{\street{Street}, \city{1-1 Namiki}, \state{Tsukuba 305-0044}, \country{Japan}}}

\affil[5]{\orgdiv{Swiss Nanoscience Institute}, \orgname{University of Basel}, \orgaddress{\city{Basel}, \postcode{4056}, \country{Switzerland}}}

%\affil[3]{\orgdiv{Department}, \orgname{Organization}, \orgaddress{\street{Street}, \city{City}, %\postcode{610101}, \state{State}, \country{Country}}}

\abstract{Twisted graphene based moir\'{e} heterostructures host a flat band at the magic
angles where the kinetic energy of the charge carriers is quenched and 
interaction effects dominate. This results in emergent phases
such as superconductors and correlated insulators that are electrostatically
tunable. We investigate superconductivity in twisted
trilayer graphene (TTG) by integrating it as the weak link in a 
superconducting quantum interference device (SQUID).
The measured current phase relation (CPR) yields a large and 
tunable kinetic inductance, up to 150~nH per square, 
of the electron and hole type intrinsic superconductors. 
We further show that the specific kinetic inductance and the critical current density 
are universally related via the superconducting coherence length, and extract
an upper bound of $\sim 200$~nm for the coherence length. Our work opens avenues for using 
graphene-based superconductors as tunable elements in superconducting circuits.
}
%phase winding
%that are electrostatically tunable


\maketitle
The discovery of novel phases, such as superconductors and correlated insulators, that emerge in a few atomic layers of twisted graphene has brought a new class of materials to the fore~\cite{bistritzer2011moire, cao2018unconventional, cao2018correlated, khalaf2019magic, lu2019superconductors, park2021tunable, hao2021electric, park2022robust}. While graphene by itself is a semi-metal, twisted graphene systems host an emergent flat band at the magic angles where the kinetic energy of the charge carriers is small and interactions dominate. Such strong electron-electron correlations lead to a spontaneous breaking of the underlying symmetries and several phases manifest from the ensuing competition between various ground states~\cite{bistritzer2011moire, cao2018unconventional, cao2018correlated, lu2019superconductors}. The superconducting phase is particularly interesting and many studies point to a departure from a Bardeen–Cooper–Schrieffer (BCS) origin of the superconductivity~\cite{cao2018unconventional, yankowitz2019tuning, lu2019superconductors, hao2021electric, kim2022evidence}. Moreover, as this rich phase diagram is accessible by electrostatic tuning, these materials have been used to design novel superconducting devices~\cite{de2021gate, rodan2021highly}.

Currently, superconducting circuits are widely utilised to design well defined quantum states and control circuits at the mesoscopic scale to create one of the leading quantum bit (qubit) architectures~\cite{vool2017introduction, krantz2019quantum, rasmussen2021superconducting}. A key fundamental property of the superconductors is called the kinetic inductance, that arises from the inertia of Cooper pairs and is proportional to the ratio of the effective mass of the charge carriers and the superfluid density.  Recently, high kinetic inductance superconductors, such as TiN~\cite{leduc2010titanium}, NbN~\cite{niepce2019high, frasca2023nbn}, NbTiN~\cite{vissers2015frequency, samkharadze2016high} and granular aluminum~\cite{rotzinger2016aluminium, maleeva2018circuit}, have been utilised in superconducting circuits, for example, to design novel qubits~\cite{grunhaupt2019granular}, to achieve a strong qubit-photon coupling~\cite{landig2018coherent, samkharadze2018strong, yu2023strong, ungerer2024strong} as well as in applications such as sensitive photon detectors~\cite{cabrera1998detection, day2003broadband, valenti2019interplay} and cryogenic amplifiers~\cite{ho2012wideband, frasca2024three}. However, the kinetic inductance of such disordered superconducting films is typically limited to $\sim2$~nH$/\square$~\cite{valenti2019interplay}. On the other hand, in moir\'{e} flat band materials a large tunable kinetic inductance is expected~\cite{portoles2022tunable}, owing to a large effective mass of the charge carriers and a tunable, low charge carrier density ($\sim10^{16}$~m$^{-2}$).

In this work, we investigate superconductivity in twisted trilayer graphene (TTG), by integrating it as the weak link in a SQUID (superconducting quantum interference devices) loop of superconducting molybdenum rhenium (MoRe). We show that, in the intrinsic superconducting regimes, the phase winds like a linear inductor across the TTG weak link and contrast this with the proximity superconducting regime where TTG exhibits a sinusoidal current phase relation (CPR). For the superconducting TTG we measure a large kinetic inductance, up to 150 nH$/\square$ which is electrostatically tunable. Moreover, by studying the kinetic inductance of the superconducting phase as a function of the critical current density we extract the upper bounds of $\approx$200~nm and $\approx$190~nm for the superconducting coherence lengths in the electron and the hole type flat bands, respectively. Finally, with a critical current density of $\sim200$~nA/$\upmu$m we show that graphene-based superconductors may be promising as tunable elements in superconducting circuits.

\begin{figure*}[t]
\captionsetup[subfigure]{labelformat=nocaption}
\includegraphics[width=2\columnwidth]{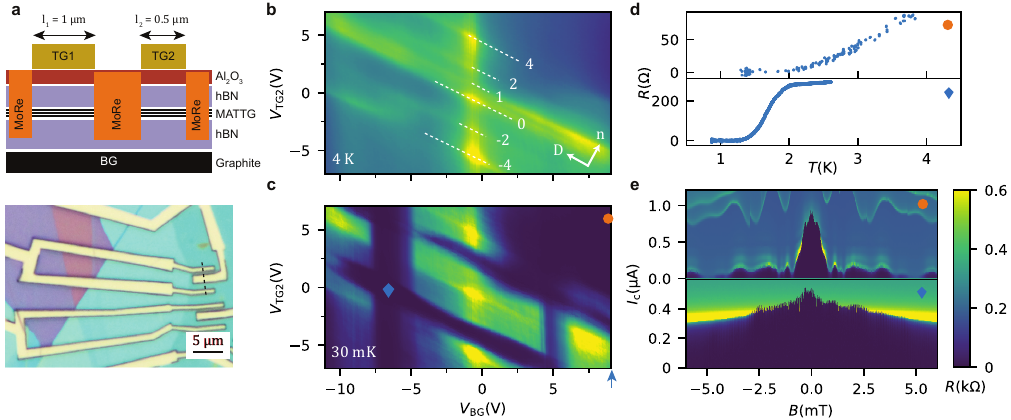}
\caption{\textbf{Device characterization: a.} The schematic (top) shows a cross-section of the device along the black dashed line in the optical image of the device (bottom), taken before the topgates are deposited.  The length ($l$) and the width ($w$) for the two junctions are $\textrm{jn1}$:  $l_{1}=1$~$\upmu$m, $w_{1}=1$~$\upmu$m and $\textrm{jn2}$: $l_{2}=0.5$~$\upmu$m, $w_{2}=2.5$~$\upmu$m. \textbf{b.} Differential resistance ($R = dV/dI$) as a function of $V_{\text{BG}}$ and $V_{\text{TG2}}$ voltage, with $V_{\text{TG1}}=0$ at $4$~K. The white dashed lines represent the charge neutrality point and the half and full filling fractions of the moir\'{e} band for $\textrm{jn2}$ ($\nu$ = 0, 1, $\pm$2 and $\pm$4). \textbf{c.} Differential resistance map at $30$~mK reveals prominent superconducting phases. \textbf{d.} $R$ \textit{vs.} Temperature ($T$) in the regions of proximity induced superconductivity (orange circle) and intrinsic superconductivity (blue diamond) in \textbf{c.} \textbf{e.} $R$ as a function of the magnetic field $B$ and the applied dc current bias $I$ in the proximity superconductor (top) and the intrinsic superconductor (bottom) regions. The color bar shown is shared with \textbf{b.} and \textbf{c.}} 
%The junctions are independently controlled by a global graphite (black) backgate (BG) as well as local Au topgates (gold) for $\textrm{J}_{1}$ ($\textrm{TG}_{1}$) and $\textrm{J}_{2}$($\textrm{TG}_{2}$).
%The TTG is encapsulated between two hBN (blue) layers, and then edge contacted by MoRe (orange) to form the SQUID junctions.
\label{Figure 1}
\begin{subfigure}{0\linewidth}
\caption{}\label{image-1a}
\end{subfigure}
\begin{subfigure}{0\linewidth}
\caption{}\label{image-1b}
\end{subfigure}
\begin{subfigure}{0\linewidth}
\caption{}\label{image-1c}
\end{subfigure}
\begin{subfigure}{0\linewidth}
\caption{}\label{image-1d}
\end{subfigure}
\begin{subfigure}{0\linewidth}
\caption{}\label{image-1e}
\end{subfigure}
\end{figure*}

We create the TTG by twisting three layers of graphene, 
such that the top and the bottom layers are aligned while
the middle layer is twisted at an angle $\approx1.7^{\circ}$. TTG is
encapsulated between hexagonal Boron Nitride (hBN) layers with graphite
underneath which functions as the backgate (BG). The schematic shown
in Fig. \ref{image-1a} (top) represents the cross section of a device along the
dashed black line in the optical image shown at the bottom. The edge
contacts to twisted trilayer graphene are made by etching the structure
and then sputtering $100$~nm MoRe. MoRe also forms the superconducting
loop for the SQUID. Finally, $25$~nm Al$_{2}$O$_{3}$ is grown
using atomic layer deposition and Ti/Au topgates are fabricated on
top of each TTG weak link for individual control over the junctions.
See methods for further details.

\begin{figure*}[th]
\captionsetup[subfigure]{labelformat=nocaption}
\includegraphics[width=2\columnwidth]{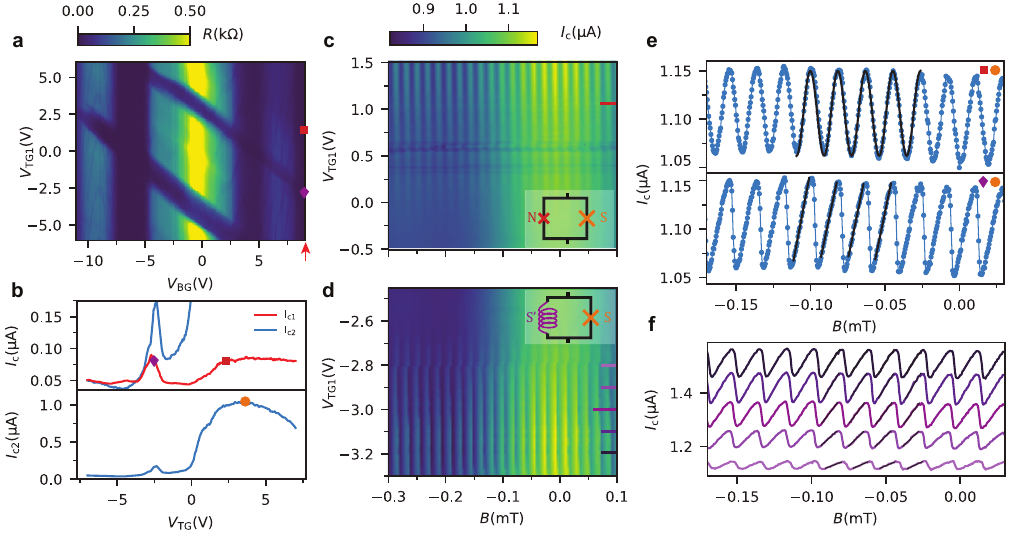}
\caption{\textbf{Current phase relation: a.} $R$ as a function of $V_{\text{BG}}$ and $V_{\text{TG1}}$, with $V_{\text{TG2}}=0$. The purple diamond (red square)  shows the intrinsic (proximity) superconductor. 
\textbf{b.} Critical current $I_{\text{c1}}$ \textit{vs.} $V_{\text{TG1}}$ is shown by the red curve (along the arrow marked in \textbf{a}) and the blue curve shows $I_{\text{c2}}$ \textit{vs.} $V_{\text{TG2}}$ (along the arrow marked in Fig.~1c), both at $V_{\text{BG}}=9$~V and $B=0$. Orange dot marks the $I_{\text{c2}}$ maximum in the proximity regime for $\textrm{jn2}$ (also shown in Fig.~1c). 
\textbf{c.} $I_{\text{c}}$ oscillations in magnetic field $B$ when operated as a SQUID with varying $V_{\text{TG1}}$, here $\textrm{jn1}$ is in the proximity regime and 
\textbf{d.} where $\textrm{jn1}$ is an intrinsic electron-type superconductor. For \textbf{c.} and \textbf{d.} $\textrm{jn2}$ is fixed to $I_{\text{c2}}^{0}\approx1$~$\upmu$A. Insets in \textbf{c.} and \textbf{d.} represent the configuration of the junctions in the SQUID.
\textbf{e.} $I_{\text{c}}$ oscillations where the upper (lower) panel shows a horizontal cut along the red (long purple) line in \textbf{c} (\textbf{d}). The black curve in the top panel is a sinusoidal fit and the black lines in the lower panel are linear fits.
\textbf{f.} $I_{\text{c}}$ oscillations in $B$ for the SQUID at various gate voltages $V_{\text{TG1}}$ when $\textrm{jn1}$ is an intrinsic superconductor, marked with the same color in \textbf{d}. The curves are vertically offset by $90$~nA for clarity. Black lines show linear fits.}
\label{Figure 2}
\begin{subfigure}{0\linewidth}
\caption{}\label{image-2a}
\end{subfigure}
\begin{subfigure}{0\linewidth}
\caption{}\label{image-2b}
\end{subfigure}
\begin{subfigure}{0\linewidth}
\caption{}\label{image-2c}
\end{subfigure}
\begin{subfigure}{0\linewidth}
\caption{}\label{image-2d}
\end{subfigure}
\begin{subfigure}{0\linewidth}
\caption{}\label{image-2e}
\end{subfigure}
\begin{subfigure}{0\linewidth}
\caption{}\label{image-2f}
\end{subfigure}
\end{figure*}

The characteristic transport signatures of twisted trilayer graphene
flat bands are shown by the differential resistance map $R=dV/dI$
at $4$~K by sweeping the backgate voltage ($V_{\text{BG}}$) and
the topgate voltage ($V_{\text{TG2}}$) on junction 2 (jn2), while $V_{\text{TG1}}=0$,
as shown in Fig.\ref{image-1b}. The dashed white lines mark the resistive features visible at integer fillings  $\nu=0,1,\pm2, 4$ of the two flat bands. Here $\nu=4n/n_{0}$, where $n$ is the charge carrier density and $n_{0}$ is the carrier density at the full filling of the flat bands. Displacement field ($D$) dependent resistance peaks are visible at $\nu=\pm4$, and yield a twist angle of $\sim1.7^{\circ}$ (Supplementary Information, section I). At $30$~mK (Fig. 1c), the zero resistance features dominate and the intrinsic superconductivity of TTG is visible on the electron and the hole side near $\nu=\pm2$
as well as the proximity superconductivity induced by MoRe at high
electron doping, in the upper right corner of Fig. 1c.

The intrinsic and proximity superconducting regions show a marked difference in the $R$ \textit{vs.} $T$ plot shown in Fig. 1d. The resistance in the proximity region (orange dot in Fig. 1c) gradually rises, highlighting the weakening proximity effect with increasing temperature. The intrinsic TTG superconductor (blue diamond in Fig. 1c), on the other hand, shows a Berezinskii–Kosterlitz–Thouless (BKT) like transition to the normal state typical of 2D superconductors~\cite{tsen2016nature, park2021tunable, hao2021electric}. Furthermore, as shown in Fig. 1e, the proximity superconductor shows a Fraunhofer-like interference in magnetic field~\cite{tinkham2004introduction}, while the intrinsic superconductor screens the external magnetic field and the critical current decays monotonically on a larger magnetic field scale~\cite{benkraouda1998critical, park2021tunable}.

The differential resistance map obtained by sweeping the backgate
voltage ($V_{\text{BG}}$) and the topgate voltage ($V_{\text{TG1}}$),
while $V_{\text{TG2}}=0$, is shown in Fig.2a. We measure the switching statistics of the superconducting junctions using a counter (for details see Supplementary Information, section II). The average switching currents for the two junctions are shown, in Fig. 2b, as a function of the respective topgate voltage at fixed $V_{\text{BG}}=9$~V (along the red arrow in Fig. 2a and
the blue arrow on Fig. 1c). We notice that
the maximum supercurrent of the intrinsic electron type superconductor,
at $V_{\text{TG1}}=V_{\text{TG2}}\approx-2.5$~V, in the two junctions
differs by a factor $\approx2-3$ (given the zero error in the
measurements). This matches well with the ratio of the widths of the
two junctions ($w_{2}/w_{1}\approx2.5$) as would be expected from
an intrinsic superconductor. On the other hand, the maximum supercurrents in the
proximity regime for the two junctions differ by more than a factor
of $10$, highlighting that the superconducting order parameter induced from the
MoRe decays exponentially with the junction lengths ($l_{1}=1$~$\upmu$m and $l_{2}=0.5$~$\upmu$m ).

The rich phase diagram for the two TTG junctions allows us to study the SQUID oscillations and probe the current phase relation (CPR) of the weak links in various configurations. The critical current for a weak link is governed by the phase drop across the junction and is given by $I_{\text{ci}}=I_{\text{ci}}^{0}f_{i}(\theta_{i})$, where $I_{\text{ci}}$ ($I_{\text{ci}}^{0}$) is the (maximum) critical current, $\theta_{i}$ is the phase difference of the superconducting order parameter across the junction $i=(1,2)$ and $f_{i}(\theta_{i})$ is its CPR, which is typically a $2\pi$ periodic function bound to $\pm1$. Since the intrinsic nature of a weak link determines its CPR, measuring the CPR is a valuable probe to access fundamental material properties~\cite{goswami2016quantum, portoles2022tunable, endres2023current, dausy2021impact}. For example, the CPR for insulating barriers is sinusoidal~\cite{tinkham2004introduction}, that for short junctions of normal materials depends on the transparency of the current carrying modes~\cite{bagwell1992suppression, golubov2004current} and a topological weak link generates a sawtooth-like CPR with $4\pi$ periodicity~\cite{beenakker2013fermion, murani2017ballistic, bernard2023long}. 

To measure the CPR of junction 1 (jn1), for example, the SQUID can be configured asymmetrically, $I_{\text{c1}}^{0}\ll I_{\text{c2}}^{0}$. This fixes the phase drop across jn2 at the critical phase (phase at which $I_{\text{c2}}$ is maximum) resulting in the external flux ($\Phi_{\text{ext}}$) essentially tuning the phase drop across jn1. However, an accurate measurement of the CPR requires careful considerations~\cite{endres2023current, babich2023limitations}. The frequently used criterion for the SQUID asymmetry, $I_{\text{c2}}^{0}/I_{\text{c1}}^{0}\sim10$ may not be sufficient and the suitable ratio depends on the CPR itself~\cite{babich2023limitations}. Moreover, the presence of large inductance effects in the SQUID loop skews the measured CPR~\cite{endres2023current}.

%e.g., the presence of large inductance effects (Supplementary Section ?) also leads to a sawtooth-like CPR that can be mistaken for topological effects~\cite{murani2017ballistic, bernard2023long}. Moreover, often the CPR of a junction is measured by configuring the SQUID asymmetrically~\cite{PhysRevLett.99.127005}, i.e. $I_{c1}\ll I_{c2}$. And typically, a ratio $I_{c2}/I_{c1}\sim10$ is considered sufficient. However, this condition depends on the CPR itself, since the necessary condition for keeping the phase of the reference junction fixed is on the derivatives of the CPR~\cite{babich2023limitations}.

For an accurate measurement of the CPR we keep the ratio $ I_{\text{c2}}^{0}/I_{\text{c1}}^{0}\geqslant\text{20}$ and ensure that we observe the expected oscillation amplitude by using the gate tunability of our weak links to eliminate the possibility of large inductance effects from the SQUID loop. We study the CPR of jn1 by fixing jn2 as the reference junction with $I_{\text{c2}}^{0}\approx1$~$\upmu$A, shown by the orange dot in Fig. 2b, and vary $V_{\text{TG1}}$ across the proximity regime of jn1 (Fig. 2c) and the intrinsic electron type superconductor (Fig. 2d) (see Supplementary Information, section IV for other configurations). The SQUID oscillations in the two regions show a stark contrast, that we better highlight in the two cuts shown in the proximity (Fig.~2e top) and the intrinsic regions (Fig.~2e bottom). For nearly the same critical current $I_{\text{c1}}^{0}\approx40$~nA, the proximitised junction shows a sinusoidal CPR, whereas the same junction when gate tuned into an intrinsic superconductor shows a highly skewed CPR. 

%We subtract the background current from jn2 from all our CPR plots unless specified otherwise.

The CPR in the proximity regime is sinusoidal ($I_{c1}=I_{c1}^{0}\sin\theta_{1}$, solid black curve in Fig.~2e) and can be understood by modeling jn1 as a short SNS junction characterised by a small effective transparency $\tau\lesssim$0.5 of the current carrying modes~\cite{bagwell1992suppression, golubov2004current} (also see Supplementary Information, section III). In contrast, the SS'S junction formed when TTG is intrinsically superconducting can be modeled as a linear inductor in a simple picture with a supercurrent $I_{s}=\Phi_{1}/L_{k}$, where $\Phi_{1}=(\theta_{1}/2\pi)\Phi_0$ and $\Phi_{0}=h/(2e)$. In addition, here the $2\pi$ periodicity of such an inductor results from the fact that at a critical phase of $\theta_{1c}\approx (2n\pm1)\pi$, where $n$ is an integer, it becomes favorable for a vortex to enter the SQUID loop rather than the phase across the intrinsic superconductor (S') winding further. In fact, vortex generation may be particularly favorable in our two-dimensional superconductor where the magnetic field screening is weak. The SQUID critical current can then be written as 

\begin{equation}
I_{c}=\frac{\Phi_{0}}{2\pi L_{k}}(\theta_{1}-2\pi n) + I_{c2}^{0}.\label{eq:linear_inductor}
\end{equation}
%where n is an integer and $I_{c1}^{0}\equiv \Phi_{0}/{2 L_{k}}$.

\begin{figure}[t]
\includegraphics[width=1\columnwidth]{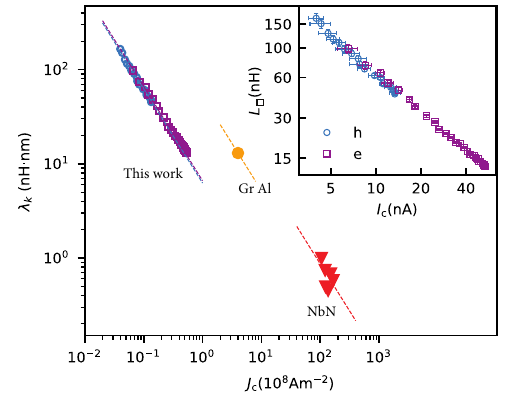}
\label{Figure 3}
\caption{\textbf{Coherence length} A plot of the specific kinetic inductance $\lambda_{k}$ vs critical current density $J_{c}$ of TTG compared to granular aluminium (Gr Al)~\cite{Thesis_Winkel2020_1000124333} and NbN~\cite{frasca2023nbn}. The dashed lines represent the least squares fit of our data to $\lambda_{k}\propto 1/J_{c}$. The inset shows the sheet kinetic inductance $L_{\square}$ vs the critical current $I_{c}$ for the hole (blue circles) and the electron (purple squares) type superconductors.}
\end{figure}

The kinetic inductance of the intrinsic superconductor (jn1) can then be extracted from the slope $\delta I_{c}/ \delta \theta_{1}$ ($\equiv \delta I_{c}/ \delta \Phi_{\text{ext}}$ for $I_{\text{c1}}^{0}\ll I_{\text{c2}}^{0}$) of the rising linear part of the measured CPR. This yields, for example, $L_{k}\approx16$~nH for the bottom panel in Fig. 2e. Multiple such oscillations are shown in Fig. 2f, as $I_{c1}^{0}$ is varied from 5~nA to 45~nA by tuning $V_{\text{TG1}}$. We see an increase in the slope of the oscillations with increasing $I_{c1}^{0}$ indicating a decreasing $L_{k}$. The extracted sheet kinetic inductance values, $L_{\square}=L_{k}w/l$, are shown as a function of $I_{c}$ for the electron (hole) type intrinsic superconductor as purple squares (blue circles) in a log-log plot in the inset of Fig.~3 (see Supplementary Information, section IV for full measurements). We emphasise that in an intrinsic TTG superconductor we are able to tune the kinetic inductance by more than an order of magnitude from 10~nH to 150~nH, with maximum kinetic inductance values of $\sim 150$~nH/$\square$ for the hole type superconductor, and $\sim 100$~nH/$\square$ for the electron type superconductor. 

Using the Ginzburg Landau (GL) theory~\cite{tinkham2004introduction, lindelof1981superconducting}, one can write the critical current density of a superconducting thin film as $J_{c}=2e\hbar/(3\sqrt{3})\times n_{s}/(m_{e}\xi)$, under the assumption that $l\geq 3.5\xi$, where $m_{e}$ is the effective mass of electrons ($2m_{e}$ is the mass of a Cooper pair), $n_{s}$ is the superfluid density, and $\xi$ is the coherence length. The kinetic inductance on the other hand can be obtained by considering that the total kinetic energy of the Cooper pairs $(n_{s}lwt) m_{e}v_{s}^{2}$ is stored in an inductor ($L_{k}I_{s}^{2}/2$), where $t$ is the film thickness and $I_{s}$ is the supercurrent. Equating the two we obtain the kinetic inductance as $L_{k}=m_{e}/(2n_{s}e^{2})\times l/(wt)$. 

We then define the specific kinetic inductance as $\lambda_{k}=L_{k}\times wt/l=m_{e}/(2n_{s}e^{2})$. This yields $\lambda_{k}=\hbar/(3\sqrt{3}e\xi J_{c})$ $-$ a universal relation between $\lambda_{k}$ and $J_{c}$ that only depends on the coherence length $\xi$  of the superconductor. We note that the contribution of jn.1 towards the critical current of the SQUID,  that we obtain from the SQUID oscillations, may be smaller than the critical current predicted from the GL theory. The former is governed by the fluxoid quantisation in the SQUID loop and the phase slip rates in the superconducting thin film, while the latter is related to the vanishing superconducting order parameter of the thin film as the velocity of superconducting charge carriers increases (see Supplementary Information, section VI).

The plot of $\lambda_{k}$ \textit{vs.} $J_{c}$, shown in Fig.~3, then allows us to extract the upper bound for $\xi$ for the hole ($\approx200$~nm) and the electron type superconductor ($\approx190$~nm) as we tune the critical current density in TTG by tuning the charge carrier density. The estimates of $J_{c}$ from the switching dynamics of an individual junction further constraint the extracted $\xi$ (see Supplementary Information, section VI). 

We emphasise that the coherence length we extract for the twisted trilayer graphene superconductor is larger than those previously reported~\cite{park2021tunable, hao2021electric} which places constraints on the nature of superconductivity. Finally, from the available literature we show two other commonly used high kinetic inductance materials, in Fig.~3, namely granular aluminum (Gr Al)~\cite{Thesis_Winkel2020_1000124333} and NbN~\cite{frasca2023nbn}, for which we extract $\xi\approx24$~nm and $\xi\approx15$~nm, respectively, in agreement with the reported values~\cite{friedrich2019onset}.

In the future, it will be intriguing to investigate the kinetic impedance at high frequencies and study other crystalline van der Waals superconductors~\cite{zhou2021superconductivity, zhou2022isospin, zhang2023enhanced}. Our results demonstrate that twisted trilayer graphene superconductors exhibit a large kinetic inductance up to $\sim 150$~nH$/\square$ that is tunable by an electric field. They also show a critical current density up to $\sim200$~nA/$\upmu$m. And while challenges in terms of the scalability of such materials are outstanding they may compliment the existing elements available for superconducting circuits. 

%\bibliography{KI}% common bib file

\section*{Methods}

\subsection*{Fabrication Details}

The graphene and hBN flakes are exfoliated on a doped silicon wafer with 285nm of silicon oxide on top (Si/SiO$_{\text{2}}$).  Large monolayer graphene flakes are selected by optical contrast under a microscope. The graphene flake selected for the device (with dimensions of around 45 $\mu$m $\times$ 15 $\mu$m) is cut into three segments using a tungsten needle controlled with a micromanipulator. For picking up the flakes, we use a polydimethylsiloxane/polycarbonate (PDMS/PC) film. We assemble the stack using a dry pick-up method ~\cite{kim2016van}. First, the top hBN (thickness = $6$~nm) flake is picked up with the stage temperature at $\sim80^{\circ}$C. The stage is then cooled down to $\sim30^{\circ}$C to pick up the first (cut) graphene segment, and subsequently rotated by $\sim1.7^{\circ}$ before picking up the second graphene segment. We revert the stage back to $\sim0^{\circ}$ to pick up the final graphene flake, such that the first and third graphene layers are aligned. This forms the alternating twist-angle trilayer graphene heterostructure. We encapsulate the TTG by picking up a bottom hBN flake (thickness = $28$~nm) at $\sim80^{\circ}$C, before finally picking up a graphite flake, (which serves as the backgate) at $\sim40^{\circ}$C. The stack is deposited by melting the PDMS/PC at $\sim180^{\circ}$C on an undoped Si/SiO$_{\text{2}}$ substrate. The PDMS/PC film residue is washed away using dichloromethane.

%The exfoliation is done using blue adhesive tape, and to ensure a higher yield of large graphene flakes, the Si/SiO$_{\text{2}}$ wafer undergoes reactive ion plasma etching for 30s right before the flakes are transferred onto it from the blue tape.

%Our twist angle of $\sim1.7^{\circ}$ is larger than the theoretically determined "magic angle" for TTG, as we anticipate some relaxation between the graphene layers due to heating in future stages of the device fabrication.

After the preparation of the stack, the regions where MoRe will be deposited as edge contacts are defined using e-beam lithography. We then etch through the top hBN, graphene and a small portion ($20\%$) of the bottom hBN above these regions using reactive ion etching (RIE) (CHF3/O2), before sputtering the MoRe edge contacts ($100$~nm). Subsequently, the graphene junction dimensions are determined by defining an etching mask in another stage of e-beam lithography, which is then etched away with CHF$_{3}$/O$_{2}$ RIE. For the next step, we deposit an additional insulating layer in the form of $25$~nm Al$_{2}$O$_{3}$ using atomic layer deposition (ALD). In the last stage of lithography, the individual togates are fabricated by defining and evaporating Ti/Au ($10$~nm/$200$~nm) above the junction weak links.

\subsection*{Measurement set-up}

The transport measurements are done in a $^{3}He - ^{4}He$ dilution refrigerator, which can be cooled down to 30 mK (base temperature). We apply a voltage signal composed of a dc component generated by a voltage source to which a small ac component supplied by a lock-in amplifier is added using a transformer. The signal is converted to a bias current using a 1 M$\Omega$ resistor connected in series with the voltage source. The device resistance is determined in a pseudo four-terminal configuration (where we can ignore line resistances but not contact resistances) by measuring the consequent voltage drop across the device on applying the bias current. The lock-in amplifier serves to reduce the signal noise and uses the small ac component (1 to 10 nA) of the signal to measure the differential resistance $R = dV/dI$.

\subsection*{Converting gate voltages to charge carrier densities}

To extract the charge carrier densities $n$ and the displacement field $D$ from the gate voltages, we utilise a parallel plate capacitor model ~\cite{Indolese2021thesis}.

\begin{equation}
n_{i} = n_{TGi} + n_{BG},
\end{equation}

\begin{equation}
D_{i} = e \frac{(n_{TGi} - n_{BG})}{2},
\end{equation}

with

\begin{equation}
n_{BG} = \frac{\epsilon_{0}\kappa_{hBN}}{d_{bottom hBN}e}(V_{BG} - V_{CNP})
\end{equation}

For the topgates, where there are two insulating layers (hBN and Al$_{2}$O$_{3}$) separating them from the graphene:

\begin{equation}
\begin{split}
n_{TGi} = \frac{1}{e}(\frac{d_{top hBN}}{\epsilon_{0}\kappa_{hBN}} + \frac{d_{Al_{2}O_{3}}}{\epsilon_{0}\kappa_{Al_{2}O_{3}}})
\\\times(V_{TGi} - V_{CNP}) 
\end{split}
\end{equation}

Here, $V_{TGi}$ is the gate voltage applied to junction $i$ by its corresponding topgate, $V_{CNP}$ is the position of the charge neutrality point, and $\kappa_{tophBN}$ ($\kappa_{Al_{2}O_{3}}$) and $d_{tophBN}$ ($d_{Al_{2}O_{3}}$) are the dielectric constants and the heights of the corresponding layers. $V_{CNP}$ can be non zero due to intrinsic doping of the sample.

\subsection*{Inductance estimation}

Our SQUID can be approximated to a rectangular loop, whose geometrical inductance is given by ~\cite{shatz2014numerical}: 

\begin{equation}
\begin{split}
L_{geo} = \frac{\mu_{0}}{\pi}((l_{1} + l_{2})\log \frac{2l_{1}l_{2}}{w+h} 
\\- l_{1}\log(l_{1} + \sqrt{l_{1}^{2} + l_{2}^{2}}) 
\\- l_{2}\log(l_{2} + \sqrt{l_{1}^{2} +l_{2}^{2}}) 
\\- \frac{l_{1}+l_{2}}{2} 
\\+ 2\sqrt{l_{1}^{2} + l_{2}^{2}} + 0.447 (h+w)),
\end{split}
\end{equation}

where $h$ is the height of the loop, $w$ is its width, and $l_{1}$ and $l_{2}$ are lengths of the two sides of the rectangular loop. We estimate $L_{geo} \approx$ 44pH. Previous studies report the sheet kinetic inductance of MoRe $L_{sq}^{MoRe}$ = 4.26~pH ~\cite{Indolese2021thesis}. This gives us the kinetic inductance of MoRe in our SQUID loop $L_{k}^{MoRe}$ = 290~pH. The total inductance of these sources is $\approx0.33$~nH, which is much smaller than the extracted loop inductance of the order 100~nH. Thus, we can assume the entire loop inductance to be contributed by the kinetic inductance of the TTG to a very good approximation. 

%\begin{@fileswfalse}
%\bibliography{KI}
%\end{@fileswfalse}


\backmatter

\bmhead{Data availability}
The nummerical data in this publication are available in numerical form at: https://doi.org/10.5281/zenodo.10732456.

%\bmhead{Supplementary information}

\bmhead{Acknowledgements}

We thank Artem Kononov and Carlo Ciaccia for fruitful discussions. This project has received funding from the European Research Council (ERC) under the European Union’s Horizon 2020 research and innovation programme: grant agreement No 787414 TopSupra, and by the Swiss Nanoscience Institute (SNI). M.B. acknowledges the support of SNSF Eccellenza grant No. PCEGP2{\_}194528, and support from the QuantERA II Programme that has received funding from the European Union’s Horizon 2020 research and innovation program under Grant Agreement No 101017733. K.W. and T.T. acknowledge support from the Elemental Strategy Initiative conducted by MEXT, Japan, the CREST (JPMJCR15F3), JST and from the JSPS KAKENHI (Grant Numbers 19H05790 and 20H00354).

%\section*{Declarations}

%\backmatter
\bmhead{Author contribution}
P.K. and C.S. designed the experiments. R.J. and P.K. fabricated the devices. R.J., M.E. and P.K. performed the measurements. R.J. performed the data analysis with inputs from P.K., C.S. and M.B. The hBN crystals were provided by K.W. and T.T.. R.J. and P.K. wrote the manuscript with inputs from all authors.

\end{document}

\maketitle

%\sectionfont{\fontsize{12}{15}\selectfont}

\section{Estimating the twist angle}

\begin{figure}
\includegraphics[width=1\columnwidth]{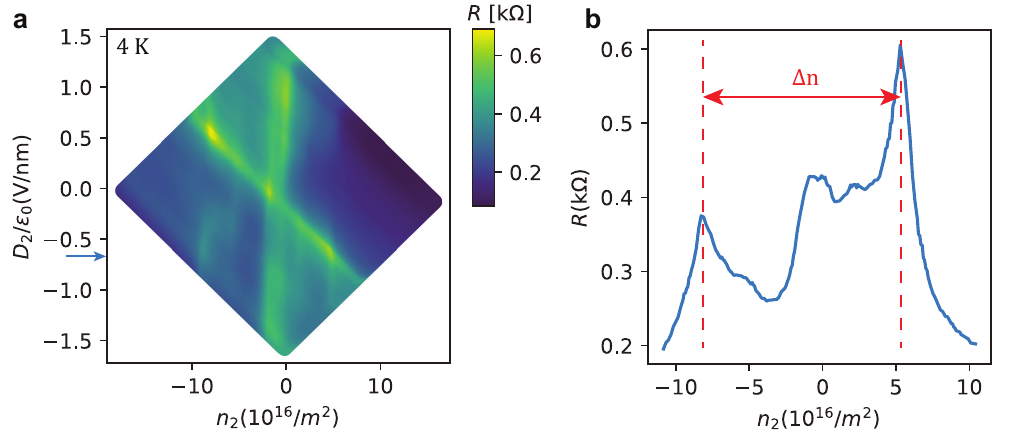}
\label{Figure S0}
\caption{\textbf{Differential resistance as a function of $n$ and $D$} \textbf{a.} $R$ as a function of $n_{\text{2}}$ and $D_{\text{2}}$ (for junction 2), taken at $T$=4K. \textbf{b.} Change in $R$ with $n_{\text{2}}$ at $D_{\text{2}}/\epsilon$=-0.6 V/nm, along the arrow in \textbf{a.} $\Delta n$=2$n_{s}$ is equal to the difference between the charge carrier densities corresponding to the full filling of the moire unit cell for the electron ($\nu$=+4) and hole side ($\nu$=-4).}
\end{figure}

To estimate the twist angle $\theta$, we use the equation ~\cite{cao2018unconventional} 

\begin{equation}
A \approx \sqrt3 a^{2}/(2\theta^{2}),
\end{equation}

where $a$ is the graphene lattice constant (=0.246 nm) and $A$ is the area of the moir\'{e} unit cell. We define $n_{s}$ as the charge carrier density corresponding to the full filling of the moir\'{e} unit cell, which gives us $n_{s}$=4/$A$, accounting for the four-fold degeneracy arising from spin and valley degeneracy. The values of $n_{s}$ for the electron and hole side can be determined from the $n$-$D$ map (Fig. S1a) with high accuracy as they correspond to resistance peaks (Fig. S1b), with $\Delta n$=2$n_{s}$ . We estimate the twist angle to be $\sim1.71^{\circ}$ for junction 1 and $\sim1.73^{\circ}$ for junction 2.

\newpage

\section{Counter measurements}

\begin{figure}[t]
\includegraphics[width=1\columnwidth]{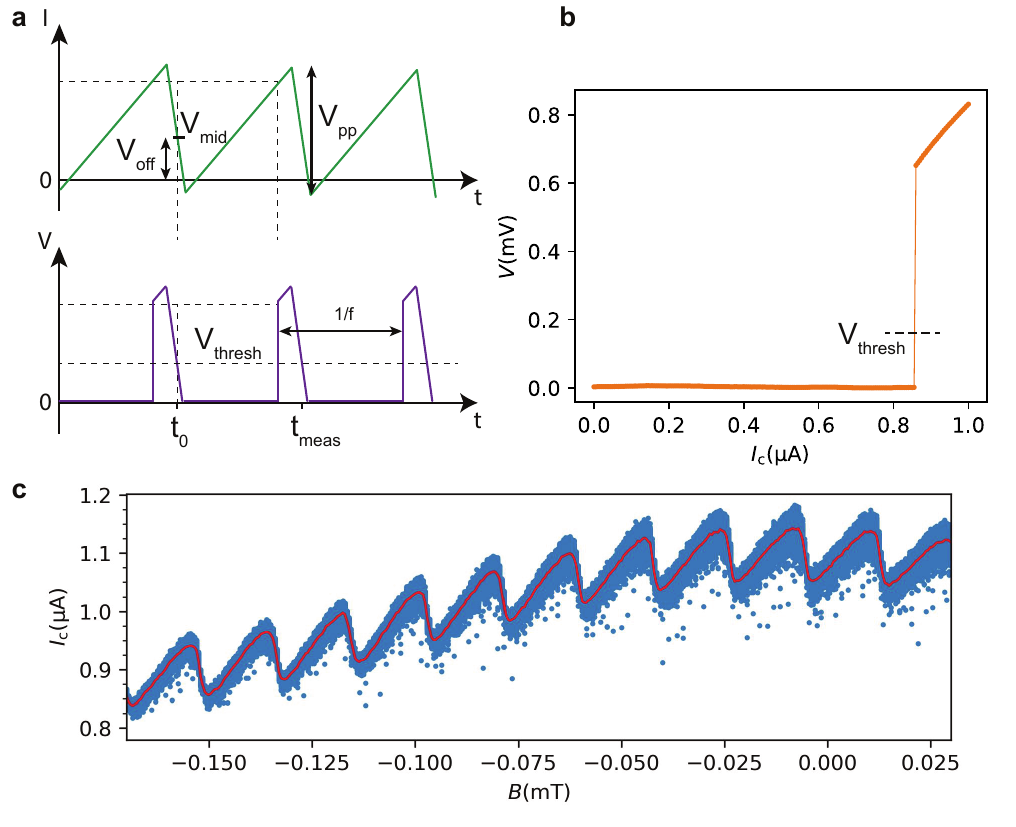}
\label{Figure S1}
\caption{\textbf{Counter measurement technique} \textbf{a.} A sawtooth shaped voltage pulse is applied in series with a resistor, creating a current pulse (top). The voltage is subsequently measured, which becomes non zero when $I>I_{c}$ (bottom). \textbf{b.} V-I (voltage-current) curve in a superconducting region of the device (shown by the orange circle in Fig. 1c of the main text). The threshold voltage $V_{thresh}$ is indicated by the black dotted line. \textbf{c.} CPR of the SQUID as measured by the counter over 200 switching events, for the electron type intrinsic superconductor (same as Fig. 2e of main text). The critical current measured over all 200 events is shown by the blue dots, while the average of the critical current is shown by the red plot. The critical current shown in the figure is plotted without subtracting the background.}
\end{figure}

In addition to experiments using the lock-in amplifier, we also perform measurements using a FCA3100 counter to measure the critical current. In this technique, a sawtooth current is generated by applying a sawtooth shaped ac voltage pulse $V_{bias}$ with frequency $f$ (277.77 Hz), amplitude $V_{pp}$ and off-set voltage $V_{off}$ from zero in series with a bias resistor $R_{bias}$ (1 M$\Omega$). The measured voltage drop across the device is then forwarded to the counter. Till the applied bias current is lesser than the critical current, there is voltage across the device. The counter is set to trigger when the measured voltage is larger than a threshold $V_{thresh}$ (100-150~mV, set above the noise level of the signal), which occurs when the bias current just exceeds the critical current, owing to a corresponding sharp rise in the measured voltage. The counter measures the time between the start time $t_{0}$ taken at 50$\%$ of $V_{bias}$ and the end time $t_{meas}$ taken when the measured voltage crosses $V_{thresh}$ (Fig. S2a). The critical current $I_{c}$ is calculated by ~\cite{endres2023current, Indolese2021thesis} 

\begin{equation}
I_{c} = \frac{1}{R_{bias}}(V_{off}+\frac{V_{ppf}}{x}(t_{meas}-\frac{1}{2f})),
\end{equation}

where $x$ denotes the ratio of the falling to the rising slope over a single period of the drive signal. Since there are large fluctuations in the switching statistics of the current, using a counter allows us to measure the current for several switching events ($N$ = 2000), and take its mean (Fig. S2c).  

We note that for an accurate measurement of the critical current, there needs to be a sharp rise in the voltage (from 0 to beyond $V_{thresh}$) at $I_{c}$ (Fig. S2b). However, this no longer remains true for small values of $I_{c}$, which introduces an offset $\approx$ 40nA in the measured critical current. However, at high values of $I_{c}$, there is a sharp rise in the voltage, which eliminates this offset. This is the regime in which we perform the switching experiments reported in this work. The large critical current of jn2 is also useful to have a large SQUID asymmetry as discussed in the main text.    

\newpage

\section{CPR fitting procedures}

\begin{figure}[t]
\includegraphics[width=1\columnwidth]{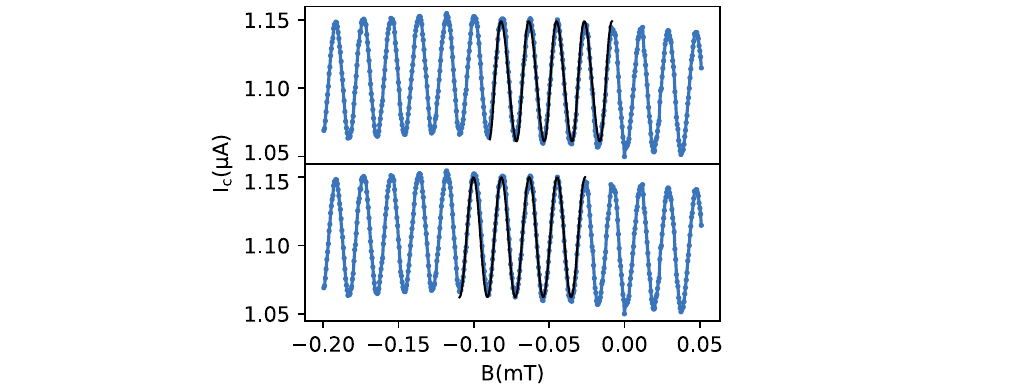}
\label{Figure S2}
\caption{\textbf{CPR fitting models} The CPR data (shown in Fig. 2e of the main text) fitted to a short junction transparency model (top) and a sinusoidal fit (bottom). The data is shown with blue dots, while the fits are shown with black lines.}
\end{figure}

For regions where both junctions have proximity induced superconductivity (SNS) (indicated by the red dot in Fig. 2a), we can assume the two CPRs to be sinusoidal, with $I_{c}$ = $I_{c1} \sin \phi_{1} + I_{c2} \sin \phi_{2}$ ~\cite{tinkham2004introduction}, where $\phi_{1}$ and $\phi_{2}$ are the phase drops across the two junctions respectively. For the highly asymmetric limit, we can assume $\phi_{2} = \pi/2$, which gives us $I_{c}$ = $I_{c2} + I_{c1} \sin \phi_{1}$. Fig. S3 (top) shows a fit of the CPR data with such a sinusoidal model, which is also shown in Fig. 2e in the main text.

Alternatively, we model $I_{c1}$ as a $2\pi$ periodic, short junction ~\cite{bagwell1992suppression}:

\begin{equation}
I_{c} = I_{c2} \left(\frac{t_{2} \sin \phi_{2}}{\sqrt {\left(1-\sin^{2}(\phi_{2}/2)\right)} }\right) + I_{c1} \left(\frac{t_{1} \sin \phi_{1}}{\sqrt {\left(1-\sin^{2}(\phi_{1}/2)\right)}}\right)
\end{equation}

where $t_{1}$ and $t_{2}$ are the transparencies of the two junctions. We plot the CPR by maximizing $I_{c}$, and find that we get the best fits to our data for $t_{1}$ = 0.54 and $t_{2}$ = 0.52 (Fig. S3 (bottom)), suggesting that both junctions are in the short diffusive limit.

\newpage

\section{Hole type superconductor characterisation}

\begin{figure}[t]
\includegraphics[width=1\columnwidth]{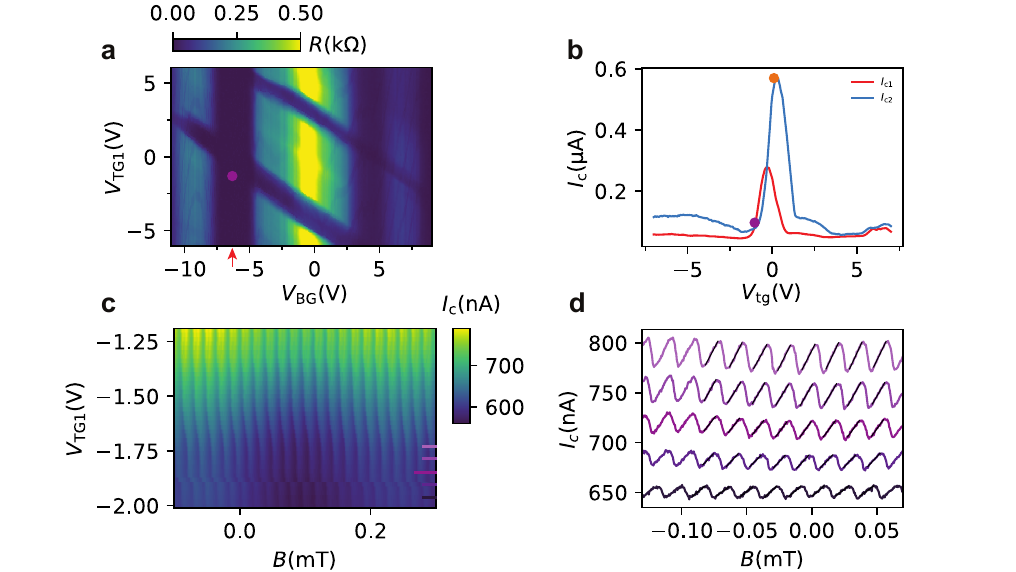}
\label{Figure S3}
\caption{\textbf{Kinetic inductance of the hole type superconductor} \textbf{a.} $R$ as a function of $V_{BG}$ and $V_{TG1}$, with $V_{TG2}$ = 0. The purple dot shows the hole type intrinsic superconducting region. \textbf{b.} Critical current vs topgate voltage measurements for both junctions. $V_{TG1}$ ($V_{TG2}$) is varied for jn 1 (jn 2) and the critical current $I_{c1}$ ($I_{c2}$) is measured, shown by the red (blue) plot. \textbf{c.} $I_{\text{c}}$ oscillations in magnetic field $B$ with varying $V_{\text{TG1}}$. \textbf{d.} CPRs at different gate voltages $V_{TG1}$, measured along the cuts shown in \textbf{c}. The black lines show the linear fits to the rising slope of the CPRs, which allows us to extract the inductance of jn1 at different $V_{TG1}$ configurations. The curves are offset by 30nA for clarity.} 
\end{figure}

In addition to distinguishing between the proximity induced and intrinsic superconductor on the electron side, we also study the CPR of the hole type superconducting region of junction 1. Fig. S4 shows the characterisation of the hole-type superconductor. Jn2 is parked at its critical current peak ($I_{c2}$=600 nA) (orange dot in Fig. S4b) and we measure the CPR of the SQUID in order to determine the kinetic inductance of jn1. While we can reach values of $I_{c1}$ up to 230 nA, we restrict our analysis to regions of $I_{c1} \leq$ 15 nA to ensure high asymmetry between the two SQUID arms as mentioned in the main text. Moreover, for $I_{c1} \geq$ 15 nA, we observe an increase in the background of the SQUID oscillations. We extract $L_{\square}$ values from 45 nH to 150 nH, with increasing slopes of the CPR (and thus decreasing $L_{\square}$) on increasing $I_{c1}$.

\newpage

\section{Extraction of coherence length}

\begin{figure}[t]
\includegraphics[width=1\columnwidth]{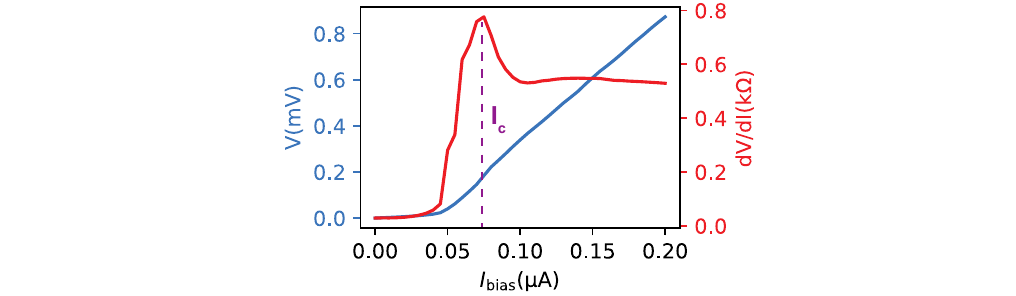}
\label{Figure S5}
\caption{\textbf{Analysis of V-I curves for jn1} \textbf{a.} $V$ as a function of the applied bias current $I_{\text{bias}}$ for the intrinsic electron type superconductor. This is taken in a configuration where jn2 is parked at the resistive state while jn 1 is in an electron type superconducting state. \textbf{b.} Differential resistance ($R = dV/dI$) as a function of the applied bias current $I_{\text{bias}}$ for the point corresponding to \textbf{a.}. The dashed line marks the bias current corresponding to the highest differential resistance.} 
\end{figure}

This section is adapted from Section 4.4 of \cite{tinkham2004introduction}, please refer to this for more details.

For a superconducting thin film of thickness much smaller than the coherence length $\xi$ (also called the Ginzburg Landau coherence length) of the superconductor, the applied bias current can perturb the complex order parameter $\psi$ from $\psi_{\infty}$, but it still remains constant everywhere. In such a system, the Ginzburg Landau equations for the current density $J_{s}$ and the free energy density $f$ can be reduced to:

\begin{equation}
J_{s} = \frac{I_{c}}{wt} = 2e|\psi|^{2}v_{s}
\end{equation}

\begin{equation}
f = f_{n} + \alpha|\psi|^{2} + \frac{\beta}{2}|\psi|^{4} + |\psi|^{2}m_{e}v_{s}^{2}+\frac{h^{2}}{8\pi}
\end{equation}

Here $m_{e}$ and $e$ are the mass and charge of an electron, $v_{s}$ is the supercurrent velocity, $h$ is the magnetic field, and $w$ and $t$ are the width and thickness of the superconducting film. 

For a thin superconducting film, the energy due to the field term $h^{2}/8\pi$ is negligible compared to the kinetic energy of the charge carriers, which allows us to ignore the last term in equation 5. For a given $v_{s}$, we can minimize the free energy density to get the optimal value of $|\psi|^{2}$. This gives us

\begin{equation}
|\psi|^{2} = \psi_{\infty}^{2} \left(1 - \frac{m_{e}v_{s}^{2}}{|\alpha|}\right)
\end{equation}

where $\psi_{\infty}^{2}=n_{s}$ is the superfluid density. Substituting this value of $|\psi|^{2}$ in equation 4, we obtain the current density

\begin{equation}
J_{s} = 2e\psi_{\infty}^{2}\left(1 - \frac{m_{e}v_{s}^{2}}{|\alpha|}\right)v_{s}
\end{equation}

Solving for $\partial J_{s}/\partial v_{s}=0$ gives us the maximum of $J_{s}$ $-$ the critical current density

\begin{equation}
J_{c} = 2e\psi_{\infty}^{2}\frac{2}{3}\left(\frac{|\alpha|}{3m_{e}}\right)^{1/2}
\end{equation}

As the Ginzburg Landau coherence length $\xi$ is given by the relation

\begin{equation}
\xi^{2} = \frac{\hbar^{2}}{4m_{e}|\alpha|}
\end{equation}

Substituting the value of $|\alpha|$ in equation 8, we get

\begin{equation}
J_{c} = \frac{2e\hbar n_{s}}{3 \sqrt 3 m_{e}  \xi}
\end{equation}

On the other hand we can determine the kinetic inductance by equating the kinetic energy of the charge carriers to the inductive energy of the wire

\begin{equation}
\frac{1}{2}(2m_{e}v_{s}^{2})(n_{s}lwt) = \frac{1}{2}L_{k}I_{s}^{2}
\end{equation}

where $l$ is the length of the superconducting film and $I_{s}$ is the supercurrent in the wire. This gives 

\begin{equation}
L_{k} = \left(\frac{m_{e}}{e^{2}n_{s}}\right)\left(\frac{l}{wt}\right)
\end{equation}

We define the sheet kinetic inductance $L_{\square}$ and the specific kinetic inductance $\lambda_{k}$ as

\begin{equation}
L_{\square} = L_{k}\left(\frac{w}{l}\right)
\end{equation}

\begin{equation}
\lambda_{k} = L_{k}\left(\frac{wt}{l}\right) = \left(\frac{m_{e}}{e^{2}n_{s}}\right)
\end{equation}

Substituting the value of $\lambda_{k}$ in equation 10, we get

\begin{equation}
\lambda_{k} = \frac{\hbar}{3\sqrt{3}e\xi}\frac{1}{J_{c}}
\end{equation}

Thus a fit to the plot of $\lambda_{k}$ \textit{vs.} $J_{c}$ allows us to extract the coherence length $\xi$.

\section{Phase slip induced switching}

As shown above, the critical current obtained from the GL theory essentially represents the maximum supercurrent (intrinsic critical current) that the thin film can sustain as the supercurrent velocity $v_{s}$ increases.
On the other hand in the SQUID geometry the oscillation amplitude is governed by the fluxoid quantisation in the SQUID loop and the phase slip rates in the superconducting thin film and the oscillation amplitude may be smaller than the intrinsic critical current. In the main text we have used the SQUID oscillation amplitude to estimate $J_{c}$. This may result in an overestimation of the coherence length.

To compare the two, we first look at the electron type superconductor in jn1 where we have measured an oscillation amplitude $(I_{c,max}-I_{c,min})/2\approx 55$~nA in the SQUID configuration, shown in main text in Fig.~3. We then analyse the V-I curves using a lock-in amplifier as the device is configured as a single SS'S junction (instead of a SQUID, by parking jn2 in its resistive state). Fig.~S5 shows one such V-I curve and the oscillation amplitude $\approx 55$~nA (mentioned above), roughly matches the span of the zero resistance state. This results in a coherence length estimate of $\approx 190$~nm (as discussed in the main text).

On the other hand, a better estimate to the intrinsic critical current may be obtained from the $I_{bias}$ value where the V-I curve deviates from its linear part (normal metal) and the vortex dynamics sets in. We estimate this intrinsic critical current by the peak in the $dV/dI$ curve, as indicated by the dashed line, which gives us $\approx75$~nA and a coherence length of $\approx 135$~nm.

A similar analysis for the hole type superconductor yields a coherence length of $\approx 50$~nm (intrinsic critical current of $\approx 56$~nA, as opposed to SQUID oscillation amplitude of $\approx 14$~nA).